\documentclass{jjap3}

\usepackage{txfonts}
\usepackage{siunitx}
\usepackage{docmute}

\title{Molecular dynamics simulation for coalescence of vacancies in tungsten crystal}
\author{Sotaro Tsuru$^{1}$\thanks{E-mail:naka-lab@nifs.ac.jp}, Hiroaki Nakamura$^{1,2}$, Yuki Goto$^{2}$, Miyuki Yajima$^{2, 3}$, Seiki Saito$^{4}$, and Shunsuke Usami$^{2}$}
\inst{$^{1}$Department of Electrical Engineering, Nagoya University, Nagoya 464-8603, Japan \\
$^{2}$National Institute for Fusion Science, National Institute of Natural Sciences, Toki 509-5292, Japan \\
$^{3}$SOKENDAI (The Graduate University for Advanced Studies), Toki 509-5292, Japan \\
$^{4}$Graduate School of Science and Engineering, Yamagata University, Yonezawa 991-8510, Japan}

\abst{We performed molecular dynamics simulations of coalescence of two vacancies in a tungsten (W) crystal to elucidate the effect of temperature and hydrogen atoms. 
Simulations were performed for two types of vacancy structures, $\mathrm{V}_9 + \mathrm{W}_1 + \mathrm{V}_9$ and $\mathrm{V}_{10} + \mathrm{W}_4 + \mathrm{V}_{10}$ ($\mathrm{V}_{n}$ means that a vacancy corresponds to the absence of $n$ W atoms, and $\mathrm{W}_{m}$ indicates that there are $m$ W atoms between two vacancies) in various cases of temperature and hydrogen atom concentration. 
Under the vacancy structure $\mathrm{V}_9 + \mathrm{W}_1 + \mathrm{V}_9$, we observed vacancy coalescence for all the cases of the temperature and the number of hydrogen atoms. 
Evaluating the potential energy required for removing one of the W atoms between two vacancies, we found that high temperature and existing hydrogen atoms in the vacancies facilitate vacancy coalescence, 
and that under the structure $\mathrm{V}_{10} + \mathrm{W}_4 + \mathrm{V}_{10}$, hydrogen atoms facilitate vacancy coalescence most strongly when the number is around 45 to 54 in each vacancy.}

\begin{document}
\maketitle

\section{Introduction}
\label{sec:introduction}

Tungsten (W) is recognized as a promising candidate for plasma-facing materials (PFMs) in future fusion reactors, 
such as ITER, due to its high melting point, low sputtering yield, and low tritium retention \cite{bolt2002, bolt2004, neu, hirai, reith}. 
As PFMs for fusion reactors, W materials are exposed to high heat fluxes by high-energy neutrons, hydrogen isotopes such as deuterium (D) and tritium (T), and helium (He) ions.
Under these conditions, W is severely damaged, and this leads to the formation of various types of defects, such as vacancies.
The defects in W are expected to lead to increased hydrogen isotope retention in W, 
because they are known to act as traps for hydrogen isotopes.\cite{fukumoto, wampler, tyburska, yajima2019, toyama, yajima2021, jin}.
High retention of hydrogen isotope in W PFMs causes various problems, 
such as embrittlement of W materials \cite{zhao, fang} and degradation of thermal conductivity \cite{cui, hu, fu}.
Therefore, it is necessary to avoid the formation of defects and high hydrogen isotope retention in W to ensure the safety and reliability of fusion reactors.

From these viewpoints, it is necessary to understand the behavior of hydrogen isotopes in W and the interaction between hydrogen isotopes and defects in W.
As a study on the interaction between hydrogen isotopes and defects in W crystal, 
Yajima et al. experimentally discovered a coalescence phenomenon of two W vacancies \cite{yajima2024}. 
In the experimental study, they found that the coalescence of vacancies occurs more easily in the case of hydrogen-containing vacancies than in the case of hydrogen-free ones. 
Furthermore, they revealed that vacancies are more likely to coalesce in higher-temperature cases.
As a major research theme for aiming to comprehensively explain the experimental results from atomic scale,
we address the clarification of the effect of hydrogen (H) content and temperatures on the coalescence of vacancies in W by using molecular dynamics (MD) simulations.
In the first step leading to the above purpose, it is required to obtain a basic understanding of the interaction between H and vacancies in W crystal.
Nakamura et al. have performed MD simulations of one vacancy in W crystal to investigate how one vacancy is affected by H atoms and temperatures \cite{nakamura}.
As their result, it has been found that more H atoms in the vacancy and higher temperatures make the vacancy structure more unstable.
This result suggests that the coalescence of vacancies occurs more easily in the presence of H atoms and at high temperatures. 
In the second step, in the present work, we perform MD simulations of coalescence of two vacancies to obtain results that will lead to the elucidation of the effect of hydrogen content and temperatures on vacancy coalescence.

In Sect. \ref{sec:method}, we describe the details of our MD simulation method.
In Sect. \ref{sec:results}, we present the results of our MD simulations.
The observation of vacancy coalescence is shown in Sect. \ref{sec:observation}, and the effect of temperature and H atoms on vacancy coalescence is discussed in Sect. \ref{sec:potential_energy}.
In Sect. \ref{sec:discussions}, we suggest underlying mechanisms to account for remarkable features found in vacancy coalescence.
Finally, in Sect. \ref{sec:conclusion}, we summarize our study.

\section{Simulation method}
\label{sec:method}

In this work, MD simulations were performed using Large-scale Atomic/Molecular Massively Parallel Simulator (LAMMPS) \cite{lammps}.
We used the embedded atom method (EAM) potential EAM1 by Bonny et al. \cite{bonny} as the interaction potential between W and H atoms.
The time step in the MD simulation was $\Delta t = 0.02$ fs, and the time in the simulation was $t = 0.0$ ps to $t = 1001.0$ ps.
The system temperature was controlled by the Langevin heat bath \cite{langevin} with a damping constant $\gamma = 0.1 \times 10^{12}\ \mathrm{s^{-1}}$ to equilibrate near the reference temperature.

\begin{figure}
  \centering
  \includegraphics[width=0.6\textwidth]{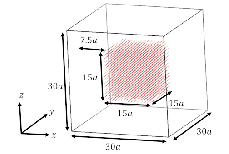}
  \caption{Initial bcc W crystal structure. The volume of the simulation box is $30a \times 30a \times 30a$, where the lattice constant $a = 3.16$ \AA, 
  and free boundary conditions are applied in all the directions. 
  The size of the W crystal is $15a \times 15a \times 15a$, and the number of W atoms is 6750.
  The W crystal is surrounded by a vacuum region with a thickness of $7.5a$.}
  \label{initW}
\end{figure}

The initial W lattice structure is shown in Fig. \ref{initW}.
At $t = 0.0\ \mathrm{ps}$, W atoms were initially arranged in a perfect body-centered cubic (bcc) crystal with a lattice constant $a$ in the center of the simulation box, where $a =  3.16$ \AA.
The volume of the crystal is $15a \times 15a \times 15a$, and the number of W atoms is 6750.
The W crystal is surrounded by a vacuum region with a thickness of $7.5a$.
The volume of the simulation box is $30a \times 30a \times 30a$, and free boundary conditions are applied in all the directions.
After the initial tungsten structure was arranged, the system reference temperature was increased from 0 K at $t = 0.0\ \mathrm{ps}$ to 563 K at $t =  0.5\ \mathrm{ps}$, and then the reference temperature is maintained at 563 K for 100 ps.

Our simulation consists of two steps: the first step is the creation of two vacancies in the W lattice, and the second step is the addition of H atoms into the vacancies.

\subsection{The first step: creation of two vacancies}

The first step is to create two vacancies in the W lattice.
We created two vacancies in the center of the W crystal by removing W atoms at $t = 100.5\ \mathrm{ps}$.
Each vacancy was created by removing $n$ W atoms from the W lattice.
Two vacancies are divided by the partition which is composed of $m$ W atoms.
Let us define the structure consisting of two vacancies and partition between them as the vacancy structure.
In what follows, we express the vacancy structure as $\mathrm{V}_n + \mathrm{W}_m + \mathrm{V}_n$.
Here $\mathrm{V}_n$ means the vacancy created by removing $n$ atoms of W, and $\mathrm{W}_m$ indicates that the partition between the two vacancies is composed of $m$ atoms of W.
After the creation of the vacancies, the reference system temperature was maintained at 563 K during 200 ps.

\subsection{The second step: addition of H atoms}

The second step is to add H atoms into the vacancies.
At the beginning of the second step, the reference system temperature was increased from 563 K at $t = 300.5\ \mathrm{ps}$ 
to the temperatures for the H irradiation at $t = 301.0\ \mathrm{ps}$.
In the experiments by Yajima et al. \cite{yajima2024}, the temperatures of the W crystal when D plasma was irradiated are 563 K and 773 K.
In this study, we chose the temperatures for the H irradiation as 563 K, 773 K, 1073 K, or 1573 K.
In what follows, the temperature for H irradiation is expressed as $T$.
After the reference system temperature was changed to $T$, it was maintained at $T$ for 100 ps.
After that, at $t = 401.0\ \mathrm{ps}$, the same number of H atoms were added into the two vacancies, inside which the H atoms were placed randomly.
The reference system temperature was maintained at $T$ to stabilize the system during 100 ps after the addition of H atoms.
At $t = 501.0$ ps, the Langevin heat bath was disconnected from the system. 
After that, simulations were performed without heat bath for 500ps.


\section{Results}
\label{sec:results}

\begin{figure}
  \centering
  \includegraphics[width=1.0\textwidth]{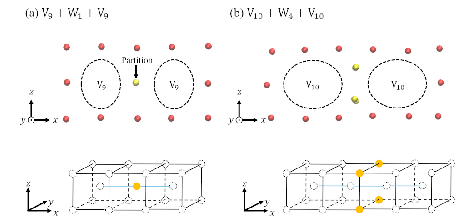}
  \caption{Two types of vacancy structures. 
  One vacancy is created by removing $n$ W atoms, and the partition between the two vacancies is composed of $m$ W atoms. 
  The structure consisting of two vacancies and partition between them is expressed as $\mathrm{V}_n + \mathrm{W}_m + \mathrm{V}_n$: (a) $\mathrm{V}_9 + \mathrm{W}_1 + \mathrm{V}_9$ structure and (b) $\mathrm{V}_{10} + \mathrm{W}_4 + \mathrm{V}_{10}$ structure.
  The projections of vacancy structures onto the x-y plane are shown in the upper parts of the figure.
  The yellow and red spheres are partition W atoms and other W atoms, respectively.
  The areas surrounded by the dashed lines are vacancies.
  The lower parts show the 3D shapes of the vacancies and the partition for the both structures.
  The dotted spheres and the yellow spheres represent the removed W atoms and the partition W atoms, respectively.
  }
  \label{Vstructure}
\end{figure}

Simulations were performed in 80 different situations for 2 patterns of the vacancy structure, 4 patterns of the system temperature $T$, and 10 patterns of the number of H atoms in one of the two vacancies $n_\mathrm{H}$.
The two patterns of vacancy structure are $\mathrm{V}_9 + \mathrm{W}_1 + \mathrm{V}_9$ and $\mathrm{V}_{10} + \mathrm{W}_4 + \mathrm{V}_{10}$.
Figure \ref{Vstructure} shows the two types of vacancy structures.
For the temperature, we chose $T= 563\ \mathrm{K}, 773\ \mathrm{K}, 1073\ \mathrm{K}, \mathrm{or}\ 1573\ \mathrm{K}$.
For the number of H atoms in one of the two vacancies, we selected $n_\mathrm{H} =  0, 9, 18, 27, 36, 45, 54, 63, 72, \mathrm{or}\ 81$.
The number of H atoms is the same for the two vacancies.
For each situation, we performed 4 simulations with different random seed numbers for the placement of H atoms.
In Sect. \ref{sec:potential_energy}, we average the quantities obtained by the 4 simulations for each situation and show the averaged values as simulation results.
In this section, we show the observation results of vacancy coalescence and quantitative analysis of the effect of $T$ and $n_\mathrm{H}$ on vacancy coalescence.

\subsection{Observation of vacancy coalescence}
\label{sec:observation}

\begin{figure}
  \centering
  \includegraphics[width=0.8\textwidth]{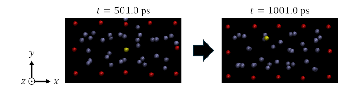}
  \caption{The vacancy coalescence in the $\mathrm{V}_9 + \mathrm{W}_1 + \mathrm{V}_9$ structure for the case of $T = 563\ \mathrm{K}$ and $n_\mathrm{H} = 18$.
  The yellow, red, and blue spheres are partition W atoms, other W atoms, and H atoms, respectively.
  The left and right pictures show the results at $t=$501.0 ps and at $t=$1001.0 ps, respectively.
  The partition W atom is in the center of the two vacancies at the beginning, and then the partition W atom moves to upper left.}
  \label{coalescence}
\end{figure}

As shown in Fig. \ref{coalescence}, we observed vacancy coalescence in the $\mathrm{V}_9 + \mathrm{W}_1 + \mathrm{V}_9$ structure.
At first, the partition W atom is in the center of the two vacancies, 
and then the partition W atom moves to upper left.
This structure change means that the two vacancies coalesce into one vacancy.
Vacancy coalescence was observed for all the cases of $T$ and $n_\mathrm{H}$ in the $\mathrm{V}_9 + \mathrm{W}_1 + \mathrm{V}_9$ structure.

In the $\mathrm{V}_{10} + \mathrm{W}_4 + \mathrm{V}_{10}$ structure, in contrast, vacancy coalescence was observed only in the case of $T = 1573\ \mathrm{K}$ and $n_\mathrm{H} = 54$, but not observed in the other cases.

We can see that vacancy coalescence is more likely to occur in the cases of 1 partition W atom than in the cases of 4 partition W atoms.
Furthermore, these results suggest that the system temperature and the number of H atoms significantly affect vacancy coalescence.

\subsection{The effect of temperature and H atoms on vacancy coalescence}
\label{sec:potential_energy}

We evaluated the effect of $T$ and $n_\mathrm{H}$ on vacancy coalescence based on the potential energy required for removing one of the partition W atoms.
However, it is difficult to obtain the potential energy for removing one W atom.
We approximated it by the potential energy difference $\Delta U(T, n_\mathrm{H})$ defined as follows:
\begin{equation}
  \Delta U(T, n_\mathrm{H}) = \frac{1}{n_{\mathrm{partition}}}\sum_{i=1}^{n_{\mathrm{partition}}} [U_{\mathrm{removed}, i}(T, n_\mathrm{H}) - U_{\mathrm{noremoved}}(T, n_\mathrm{H})],
\end{equation}
where $U_{\mathrm{removed}, i}$ is the potential energy of the system in which the $i$-th partition W atom does not exist,
and $U_{\mathrm{noremoved}}$ is the potential energy of the system in which all the partition W atoms exist.
The number of partition W atoms is $n_{\mathrm{partition}} = 1 \,\mathrm{and}\, 4$ for the $\mathrm{V}_9 + \mathrm{W}_1 + \mathrm{V}_9$ and $\mathrm{V}_{10} + \mathrm{W}_4 + \mathrm{V}_{10}$ structures, respectively.

\begin{figure}
  \centering
  \includegraphics[width=1.0\textwidth]{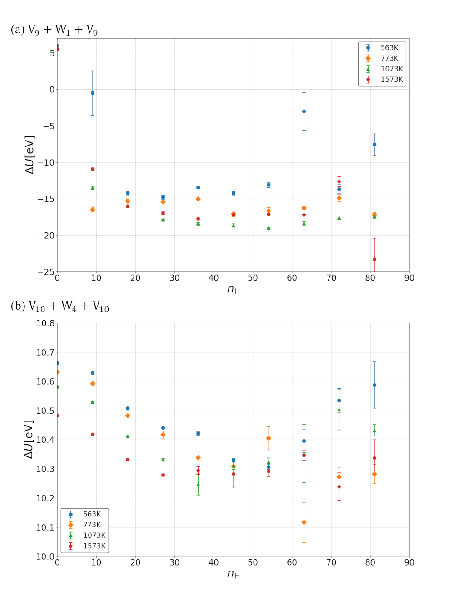}
  \caption{The potential energy difference $\Delta U(T, n_\mathrm{H})$ for (a) $\mathrm{V}_9 + \mathrm{W}_1 + \mathrm{V}_9$ and (b) $\mathrm{V}_{10} + \mathrm{W}_4 + \mathrm{V}_{10}$ structures.
  The error bars are the standard error of $\Delta U(T, n_\mathrm{H})$.
  The potential energy difference is less than 0 for all the cases $T$ and $n_\mathrm{H}$ except for $n_\mathrm{H} = 0$ in the $\mathrm{V}_9 + \mathrm{W}_1 + \mathrm{V}_9$ structure.
  In contrast, the potential energy difference is minimized around $45 \leq n_\mathrm{H} \leq 54$ in the $\mathrm{V}_{10} + \mathrm{W}_4 + \mathrm{V}_{10}$ structure.}
  \label{deltaU}
\end{figure}

As shown in Fig. \ref{deltaU}, $\Delta U(T, n_\mathrm{H})$ for the $\mathrm{V}_9 + \mathrm{W}_1 + \mathrm{V}_9$ structure is less than that for the $\mathrm{V}_{10} + \mathrm{W}_4 + \mathrm{V}_{10}$ structure.
This result indicates that vacancy coalescence is more likely to occur in the $\mathrm{V}_9 + \mathrm{W}_1 + \mathrm{V}_9$ structure than in the $\mathrm{V}_{10} + \mathrm{W}_4 + \mathrm{V}_{10}$ structure.
This supports that vacancy coalescence is observed in the $\mathrm{V}_9 + \mathrm{W}_1 + \mathrm{V}_9$ structure for all the cases of $T$ and $n_\mathrm{H}$, but not observed in the $\mathrm{V}_{10} + \mathrm{W}_4 + \mathrm{V}_{10}$ structure for most cases.

In the $\mathrm{V}_9 + \mathrm{W}_1 + \mathrm{V}_9$ structure and in the $\mathrm{V}_{10} + \mathrm{W}_4 + \mathrm{V}_{10}$ structure, 
$\Delta U(T, n_\mathrm{H}) < \Delta U(T, 0)$ holds for each case of $T= 563\ \mathrm{K}, 773\ \mathrm{K}, 1073\ \mathrm{K}, \mathrm{and}\ 1573\ \mathrm{K}$.
This result shows that the existing of H atoms reduces the robustness of the partition between the two vacancies, triggering vacancy coalescence.
In other words, H atoms facilitate vacancy coalescence.
In addition, $\Delta U(T, n_\mathrm{H})$ tends to decrease as $T$ increases.
Higher temperatures cause the partition W atom to migrate more easily, facilitating vacancy coalescence.
These results are consistent with the experimental results on the influence of D plasma radiation and high temperature \cite{yajima2024}.

In the $\mathrm{V}_{10} + \mathrm{W}_4 + \mathrm{V}_{10}$ structure, $\Delta U(T, n_\mathrm{H})$ decreases as $n_\mathrm{H}$ increases when $n_\mathrm{H} \leq 45$.
In contrast, $\Delta U(T, n_\mathrm{H})$ tends to increase as $n_\mathrm{H}$ increases when $n_\mathrm{H} \geq 54$.
This result indicates that the facilitation of vacancy coalescence depends on the number of H atoms in vacancies and is maximized around $45 \leq n_\mathrm{H} \leq 54$.

\section{Discussions}
\label{sec:discussions}

\begin{figure}
  \centering
  \includegraphics[width=1.0\textwidth]{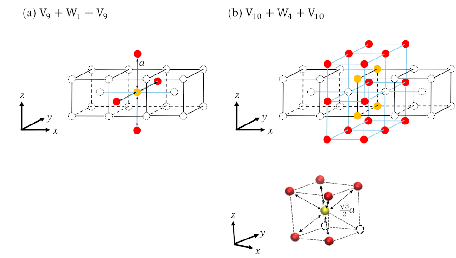}
  \caption{
  3D vacancy structures of (a) $\mathrm{V}_9 + \mathrm{W}_1 + \mathrm{V}_9$ and (b) $\mathrm{V}_{10} + \mathrm{W}_4 + \mathrm{V}_{10}$.
  The dotted spheres represent the removed W atoms.
  The yellow and red spheres represent the partition W atoms and the nearest W atoms outside the partition, respectively.
  The lower part of (b) shows one of the partition W atoms and the nearest W atoms for it in the $\mathrm{V}_{10} + \mathrm{W}_4 + \mathrm{V}_{10}$ structure.
  In the $\mathrm{V}_9 + \mathrm{W}_1 + \mathrm{V}_9$ structure, the partition W atom is surrounded by the 4 nearest W atoms at a distance of the lattice constant $a$.
  In the $\mathrm{V}_{10} + \mathrm{W}_4 + \mathrm{V}_{10}$ structure, the partition W atom is surrounded by the 6 nearest W atoms at a distance of $\sqrt{3}/2\ a$.
  }
  \label{nearest}
\end{figure}

In this study, we found that vacancy coalescence occurs easily in the $\mathrm{V}_9 + \mathrm{W}_1 + \mathrm{V}_9$ structure.
However, vacancy coalescence was rarely observed in the $\mathrm{V}_{10} + \mathrm{W}_4 + \mathrm{V}_{10}$ structure.
This difference in the vacancy coalescence is due to the distance between the partition W atoms and is owing to the nearest W atoms outside the partition and the number of the nearest W atoms. 
As shown in Fig. \ref{nearest}, in the $\mathrm{V}_9 + \mathrm{W}_1 + \mathrm{V}_9$ structure, the partition W atom is located at a distance of $a$ from the nearest W atoms, where $a$ is the lattice constant of tungsten, and the number of the nearest W atoms is 4.
On the other hand, in the $\mathrm{V}_{10} + \mathrm{W}_4 + \mathrm{V}_{10}$ structure, each partition W atom is located at a distance of $\sqrt{3}/2\ a$ from the nearest W atoms, and the number of the nearest W atoms is 6. 
According to EAM1 by Bonny et al. \cite{bonny}, in the range of $r \geq 0.76a$, where $r$ is the distance between two W atoms, the smaller $r$ is, the greater the interaction between the W atoms is.
In addition, the interaction between W atoms increases with the number of the nearest W atoms outside the partition.
Therefore, in the $\mathrm{V}_9 + \mathrm{W}_1 + \mathrm{V}_9$ structure, the interaction between the partition W atom and other W atoms outside the partition is weaker, and thus the partition W atom is easy to move from the initial position, 
while in the $\mathrm{V}_{10} + \mathrm{W}_4 + \mathrm{V}_{10}$ structure, the interaction is stronger and hence the partition W atoms are hard to move.

From Fig. \ref{deltaU}, we found that $\Delta U(T, n_\mathrm{H})$ is less than $\Delta U(T, 0)$ for each case of the temperature $T$.
This means that H atoms facilitate vacancy coalescence. 
Meanwhile, we recognize that the facilitation strength depends on $n_\mathrm{H}$.
Figure \ref{deltaU} shows that the potential energy difference is minimized around $45 \leq n_\mathrm{H} \leq 54$ in the $\mathrm{V}_{10} + \mathrm{W}_4 + \mathrm{V}_{10}$ structure.
From this result, we conclude that when the number of H atoms existing in a vacancy exceeds a threshold, which is between 45 and 54, the effect of a newly added H atom on vacancies is reversed from destabilization to stabilization. 
The former and the latter correspond to decreasing and increasing $\Delta U(T, n_\mathrm{H})$, respectively. 
This conclusion is supported by the previous studies suggesting two kinds of properties of H atoms in vacancy coalescence \cite{middleburgh, kato2015, bukonte, huang}. 
The first kind of the suggested properties plays a role in reducing the vacancy formation energy by attractive interaction between H atoms and vacancies, which means destabilizing vacancy structure \cite{middleburgh, kato2015, bukonte, huang}. 
The second kind plays a role in reducing the overall strain on the lattice by filling a vacancy and stabilizing vacancy structure \cite{middleburgh}.
%
%
%
The previous studies discussed the two properties separately and did not consider the dependence of the two properties of H atoms on the H concentration in vacancies.
In our work, we claim that the two properties depend on the H concentration in vacancies and suggest that the number of H atoms in vacancies determines which of the two properties is greater.



Furthermore, we suggest that in terms of the hydrogen filling rate, approximately 50 \% is the threshold for whether the effect of a newly added H atom is destabilization or stabilization
The previous research by Ohsawa et al. \cite{ohsawa} showed that $\mathrm{V}_1$, i.e., one vacancy created by removing one W atom, traps a maximum of 12 H atoms, and they are located at 12 T-sites of $\mathrm{V}_1$.
Eliminating overlapped sites, $\mathrm{V}_{10}$ has 94 T-sites, and thus it can trap a maximum of 94 H atoms.
Therefore, the hydrogen filling rate of 50 \% of $\mathrm{V}_{10}$ corresponds to the number of H atoms $n_\mathrm{H} = 47$ and this value is in the range of $45 \leq n_\mathrm{H} \leq 54$.
We consider that the hydrogen effect reverses, when the hydrogen filling rate in the vacancy exceeds approximately 50 \%.

\section{Conclusion}
\label{sec:conclusion}

In this study, we performed MD simulations under various conditions to investigate the effect of temperature and H atoms on the coalescence of vacancies in W crystal.
For the temperature, we chose $T= 563\ \mathrm{K}, 773\ \mathrm{K}, 1073\ \mathrm{K}, \mathrm{or}\ 1573\ \mathrm{K}$.
For the number of H atoms, we selected $n_\mathrm{H} =  0, 9, 18, 27, 36, 45, 54, 63, 72, \mathrm{or}\ 81$.
We observed the coalescence of vacancies for all the cases of $T$ and $n_\mathrm{H}$ in the $\mathrm{V}_9 + \mathrm{W}_1 + \mathrm{V}_9$ structure.
In contrast, the coalescence of vacancies was observed only in the case of $T = 1573\ \mathrm{K}$ and $n_\mathrm{H} = 54$ in the $\mathrm{V}_{10} + \mathrm{W}_4 + \mathrm{V}_{10}$ structure.
Vacancy coalescence is more likely to occur in the cases of one partition W atom ($\mathrm{W}_1$) than in the cases of four partition W atoms ($\mathrm{W}_4$).
In addition, the ease of vacancy coalescence is strongly influenced by the system temperature and the number of H atoms.
The quantitative analysis of the potential energy difference $\Delta U(T, n_\mathrm{H})$ shows that high temperature and H atoms in vacancies facilitate the coalescence of vacancies.
These results are consistent with the experimental results by Yajima et al. \cite{yajima2024} that the coalescence of vacancies was observed at high temperature and high H concentration in W crystal.
Furthermore, in the $\mathrm{V}_{10} + \mathrm{W}_4 + \mathrm{V}_{10}$ structure, we have found that the facilitation of vacancy coalescence by H atoms is maximized at $45 \leq n_\mathrm{H} \leq 54$.
This maximization of the facilitation in this range means that the effect of a newly added H atom on the vacancies changes from destabilization to stabilization, as the H concentration in the vacancies increases.

\acknowledgments
The computation was performed using Research Center for Computational Science, Okazaki, Japan (Project: 24-IMS-C099) 
and Plasma Simulator of NIFS. 
The research was supported by Japan Society for the Promotion of Science (Grant Nos. JP22H05131, JP23H04609, JP22K18272, JP23K03362, JP19K14692), 
by the NINS program of Promoting Research by Networking among Institutions (01422301), 
by the NIFS Collaborative Research Programs (NIFS22KIIP003, NIFS24KIIT009, NIFS24KIPT013, NIFS22KIGS002, NIFS22KIEP002) 
and by the Ex-CELLS Special Collaboration Program of Exploratory Research Center on Life and Living Systems (24-S5), 
and the Inter-University Cooperative Research Program of the Institute for Materials Research, Tohoku University (Proposal No. 202312-IRKMA-0514).

\end{document}